# How statistical model development can obscure inequities in STEM student outcomes

Ben Van Dusen[1] & Jayson Nissen[2]

[1]Iowa State University, School of Education, 1620 Lagomarcino Hall, 901 Stange Road, Ames, IA 50011-1041, bvd@iastate.edu
[2]Nissen Education Research and Design, 1111 Maine Ave, Slidell, LA 70458, jayson.nissen@gmail.com

## 1.   Abstract

Researchers often frame quantitative research as objective, but every step in data collection and analysis can bias findings in often unexamined ways. In this investigation, we examined how the process of selecting variables to include in regression models (model specification) can bias findings about inequities in science and math student outcomes. We identified the four most used methods for model specification in discipline-based education research about equity: a priori, statistical significance, variance explained, and information criterion. Using a quantitative critical perspective that blends statistical theory with critical theory, we reanalyzed the data from a prior publication using each of the four methods and compared the findings from each. We concluded that using information criterion produced models that best aligned with our quantitative critical perspective's emphasis on intersectionality and models with more accurate coefficients and uncertainties. Based on these findings, we recommend researchers use information criterion for specifying models about inequities in STEM student outcomes.

## 1. Introduction

Researchers and policymakers often view quantitative research as objective and fact-based, particularly in comparison to qualitative research (Stage, 2007). We see evidence of this perspective in the prevalence of quantitative analysis driving education policy (Carver, 1975; Darling-Hammond et al., 2007; Gillborn et al., 2018). Biases, however, influence quantitative findings throughout the process of data collection, analyses, and interpretation (Stage, 2007). Researchers rarely interrogate their methods and regularly ignore their potential to bias findings.

For example, mortgage lenders cannot use race when reviewing a loan application, but location information can perpetuate racist oppression in housing through redlining (Rothstein, 2017; O'Neil, 2016). These unexamined biases are particularly problematic in equity research given social science statistics' roots in the eugenics movement of the late 19th and early 20th century (Zuberi, 2001; Zuberi & Bonilla-Silva, 2008). The eugenics movement used the guise of science and objectivity to advance the racist and classist ideas of white-superiority (McGee, 2020).

This investigation examined how model specification practices in developing regression models of inequities in student outcomes can influence findings on STEM student equity. Model specification identifies a single "best" model from all of the possible models. We begin by explaining our conceptual framework, which uses a quantitative critical framework (QuantCrit; Stage, 2007; Lopez et al., 2018) to blend critical theory (Ladson-Billings, 2013; West, 1995) with statistical theory. Then, we use our conceptual framework to identify model specification goals in discipline-based education research (DBER) investigations of equity. Finally, using these goals to evaluate four model specification methods commonly used in DBER allows us to identify a preferred method. To evaluate these four model specification methods, we reanalyze data from a prior publication on equity in DBER (Van Dusen & Nissen, 2020) using each method.

To aid readers, we have included a list of statistical modeling terms in Table 1.

## 1.1 Objectives

This article demonstrates how subjective decisions in building quantitative models shape the results those models produce. To this end, we use a QuantCrit framework to identify goals for model specification in equity research, evaluate how well methods commonly used in DBER achieve those goals, and illustrate how these methods shape findings. Research questions should drive the analytical methods used in an investigation (Ding, 2019; Schmueli, 2010). In this article, we focus on work designed to describe or explain trends in existing datasets, rather than work designed to make predictions about future outcomes (Shmueli, 2010).

1. *Use critical theory to identify model specification goals in DBER investigations of equity* - Because critical theory rejects the objectivity of quantitative research, it provides a framework for identifying the model specification tensions and goals in DBER investigation of equity.

2. *Evaluate how well standard DBER model specification methods met the goals of model specification identified in our first objective* - To identify standard DBER model specification methods and the statistical processes that underpin them, we will review studies from a cross section of DBER journals. We will evaluate each method's efficacy to meet the identified goals of model specification for DBER investigations of equity.

3. *Illustrate how model specification methods can impact findings in DBER investigations of equity* - We will reanalyze the data from our prior study of equity in physics student learning (Van Dusen & Nissen, 2020) using the standard model specification methods. Comparing and contrasting the findings from each method will demonstrate how model specification practices can bias findings.

**Table 1.** Statistical modeling terms.

| Term | Definition |
|---|---|
| Model | An equation that describes a relationship between variables. |
| Linear regression | A linear equation that represents the relationship between an outcome variable and predictor variable(s). |
| Hierarchical linear regression | A linear regression method that accounts for the nested (i.e., hierarchical) nature of the data (e.g., students nested in courses). |
| Outcome/dependent variable | The variable representing the measured outcome (e.g., student learning, graduation rates, or attitudes). |
| Predictor/ independent variable | The variables that may correlate with the outcome variable that the model accounts for or measures the impact of (e.g., race, gender, treatment/control, or ACT scores). |
| Coefficient | Each predictor variable is assigned a coefficient that estimates its relationship to the outcome variable |
| Standard error | A measure of an estimate's precision/uncertainty that is often used to create compatibility intervals and calculate p-values. |
| Compatibility intervals (i.e., confidence intervals) | A measure of an estimate's precision/uncertainty. Often mistakenly interpreted as meaning that there's a 95% probability that the interval contains the true value. (Amrhein, Trafimow & Greenland, 2019) |
| p-value | The probability of obtaining test results at least as extreme as the observed results, under the assumption that the null hypothesis is correct, $P(\text{observation}|H_0)$. Does not inform the odds that the hypothesis is correct, $P(H_0|\text{observation})$ (Nuzzo, 2014). |
| Variance | A measure of the spread of the data around the predicted values from the model in squared units. For the simplest model, this is the spread around the mean. |
| Standard deviation | The square root of the variance. It measures the spread of the data in the same units as the data. |
| Variance explained (R) | The percentage decrease in the variance with the inclusion of predictor variables, often measured by $R^2$. |
| Adjusted $R^2$ | A measure of variance explained with an adjustment for additional predictor variables. The adjusted R-squared increases only if the new variable improves the model more than would be expected by chance. |
| Additional variance explained | The change in $R^2$ associated with the addition of a predictor variable. This is similar to but distinct from partial variance explained. |
| Model misspecification | Models that either include non-explanatory variables (over-specification) or exclude important variables (under-specification) and generate biased estimates and standard errors. |
| a priori | Reasoning from theoretical deduction rather than from observation. |

## 2. Author positionalities

The following is the first author's positionality statement: I identify as a White[1] cisgender, heterosexual man. I was raised in a pair of lower-income households, but I now earn an upper-middle-class income. I hold an undergraduate degree in physics and a Ph.D. in education. I was an assistant professor at a teaching-intensive, Hispanic serving institution but now am an assistant professor at a research-intensive, predominantly White institution. My experiences working with minoritized[2] students, particularly the Latino and Latina students I have had the honor to mentor as learning assistants (Otero, 2015) and as researchers, have motivated me to use my position and privilege to dismantle oppressive power structures. As someone who seeks to serve as a co-conspirator[3], it is easy to overlook my privileges. I try to broaden my perspective through feedback from those with more diverse lived experiences than my own.

The following is the second author's positionality statement: Identifying as a White, cisgendered, heterosexual, able-bodied man provides me with opportunities denied to others in American society. These identities predominate other people's initial perceptions of me and privilege me in science spaces. My experience growing up in a poor home and as a veteran of the all-male submarine service motivate me to reflect on and work to dismantle my privilege. The juxtaposition between who I am perceived as and how I perceive myself has motivated this work on building quantitative methods that respect and reflect students' identities and societal power structures. Because I am not a woman or a person of color, I brought a limited perspective to this work on racism and sexism.

To broaden the paper's perspective beyond those of the authors, we contracted McKensie Mack of Radical Copy to perform an equity audit of the manuscript.

## 3. Conceptual framework

### 3.1 Critical theory

Critical Race Theory (CRT) began in the 1970s and 1980s to address social injustices and racial oppression (Ladson-Billings, 2013; West, 1995; Sleeter & Bernal, 2004). Subsequently, scholars in many fields, including education (Ladson-Billings, 1998; Ladson-Billings, 2009), have used CRT to guide their work in areas such as LatCrit, FemCrit, AsianCrit, and WhiteCrit (Solorzano & Yosso, 2001). Each of these branches applies the defining characteristics of CRT (e.g., examining oppressive power structures, challenging the ideas of objectivity, and considering intersectionality of individual's identities; Ladson-Billings, 2013) in novel contexts.

---

[1] In this publication, we capitalize all races, including White, emphasizing that there is no default race and that they are all social constructs with associated sets of cultural practices.
[2] We use the term minoritized to reflect that students are categorized as minority through an active social process, rather than a characteristic of the individual.
[3] I define co-conspirator as someone who uses their privilege to take action against racism regardless of personal consequences.

The tenets of CRT are not explicitly qualitative, but CRT research has historically used qualitative methods. The predominance of qualitative methods in CRT investigations is, in part, due to CRT's focus on individual's narratives and counter-narratives (McGee, 2020). However, we will use a quantitative critical (QauntCrit; Stage, 2007) perspective to inform our use of quantitative methods in ways consistent with the tenets of CRT. In this investigation, we forefront three tenets of QuantCrit proposed by Covarrubias et al. (2018) and Gillborn et al. (2018): 1) disrupting dominant quantitative methods, 2), data cannot 'speak for itself' and 3) taking an intersectional perspective.

## 3.1.1 Disrupting dominant quantitative methods

Quantitative research has a long history of perpetuating inequities (Zuberi & Bonilla-Silva, 2008). The eugenics movement formed the foundation of many statistical methods in social sciences (Zuberi, 2001). The eugenics movement appropriated the scientific ideas of genetics and evolution to advocate for selective breeding to remove so-called inferior characteristics. These racist, classist, and ableist ideas used science to harm minoritized individuals and communities. U.S. states began enacting laws in 1907 that legalized sterilizing the "feebleminded" (Reilly, 2015). This has led to the coercive sterilization of over 60,000 poor, unwed, Black, Indigenous, people of color, or mentally disabled people. Coercive sterilization is still legal and occurring in the U.S. today (Wilson, 2018), for example in for-profit immigration detention centers (Jankowski, 2020). We also see the eugenics movement's ideas in modern academia, which often excludes minoritized individuals' contributions to science (McGee, 2020).

Quantitative social science research is rife with 'hidden assumptions' and 'ideological inscriptions' (Stage 2007, pg. 9) that promote "color-blind" interpretations of data. Being "color-blind" is often presented as a virtue as it is seen as race-neutral (Bonilla-Silva, 2006). Ignoring race, however, ignores a major component of individual's identities and lived experiences and obscures inequities. In quantitative analyses, race-neutral models often minimize or replace race with proxies that appear as if race does not matter. For example, the Fair Housing Act of 1968 prohibited the practice of denying mortgages to Black people based on their skin color (i.e., redlining). Still, it did not end the denial of mortgages to Black people. Banks propagated the practice by using proxies for race, such as zip-code and neighborhood median income. Some researchers have termed this practice algorithmic redlining (Allen, 2019). The racist history of race-neutral interpretations of data in conjunction with quantitative researcher's objectivity claims has mostly kept CRT researchers from implementing quantitative methods (Covarrubias et al., 2018). As QuantCrit researchers, it is incumbent upon us to disrupt oppressive systems by identifying the hidden assumptions in our work and using quantitative methods to create more just systems.

Researchers often skip rigorous model specification in favor of replicating the models from prior investigations. Relying on models from previous research to accurately describe a new dataset creates opportunities to perpetuate hidden assumptions. For example, examining the impacts of racism on student outcomes with a limited sample size may necessitate that researchers aggregate students into only two groups. STEM education researchers most often define these groups using the National Science Foundation's definition of underrepresented minorities: students who are not White or Asian unless they are Hispanic. These aggregations get replicated in future studies with large samples with the statistical power to disaggregate the data. This leads to a body of literature that fails to address the variation in how racist power structures impact social identity (e.g., gender and race) groups and perpetuates the model minority myth that Asians do not experience racism (Lee, 2015).

### 3.1.2 Data cannot 'speak for itself'

Assumptions that serve the dominant perspective can shape every stage of collecting, analyzing, and interpreting data (Covarrubias et al., 2018; Zuberi & Bonilla-Silva, 2008). To properly contextualize findings, researchers must critically examine and communicate how assumptions have influenced their data collection and analysis. The practice of reporting data and letting it "speak for itself" encourages readers to assume that the findings are objective and that they should interpret them from the dominant perspective, which reinforces existing power structures. In investigations of equity in STEM student outcomes, this practice often leads readers to ignore bias in assessment instruments and administration practices and reinforces the preexisting idea that the "achievement gap" describes inherent deficiencies in minoritized students. We strive to contextualize our findings and prevent them from supporting deficit views of students. Deficit views position differences across social identifier (e.g., gender and race) groups as arising from individuals' deficiencies rather than because of the hegemonic power structures the individuals are embedded within. To prevent our findings from being used to support deficit perspectives, we discuss the limitations of our data, present our data and methods transparently, and discuss inequities in outcomes across social identifier groups as the products of racism and sexism.

### 3.1.3 Taking an intersectional perspective

QuantCrit research assumes that differences in outcomes across social identifier groups result from intersecting oppressive power-structures. Critical research uses intersectionality (Crenshaw, 1990) to account for oppressive power structures never operating independently (Bruning et al., 2015; Armstrong & Jovanovic, 2015). For example, Black women may not experience racism in the same ways as Black men nor sexism in the same ways as White women. An intersectional approach to modeling includes interaction terms between student social identifier variables (e.g., race and gender) to account for interactions across axes of oppression (Schudde, 2018; Lopez et al., 2018). Ultimately, researchers may remove interaction terms if they are not predictive, but the model specification process must allow for intersectionality.

## 3.2 Operationalizing equity

We operationalized equity (Rodriguez et al., 2012; Stage, 2007) as *equality of learning.* Equality of learning is achieved when students from different social identifier groups learn equivalent amounts. This perspective has been called "equity for equal potential" (Espinoza, 2007) and is related to "equity of fairness" (Lee, 1999; Rodriguez et al., 2012). Equality of learning ensures that students who start the semester similarly prepared attain the same level of achievement. We use equality of learning as it is a commonly used type of equity and aligns with Van Dusen & Nissen's (2020) use of gain as the outcome variable. Still, it is a problematic form of equity because it ignores prior differences that represent the educational debts (Figure 1; Ladson-Billings, 2006) society owes to students from minoritized groups. Achieving equality of learning maintains society's educational debts. Exceeding it begins to repay those debts and failing to meet it adds to those debts. We advocate for the use of multiple and stronger anti-racist operationalizations of equity, such as equality of outcomes (sometimes called equity of parity; Rodriguez et al., 2012). Equality of outcomes is met when educational debts are fully repaid, and all groups have the same average outcomes regardless of prior educational debts. In this analysis, however, for brevity's sake we only used one operationalization of equity, equality of learning, as it was the one modeled using gain as the outcome variable by Van Dusen & Nissen (2020).

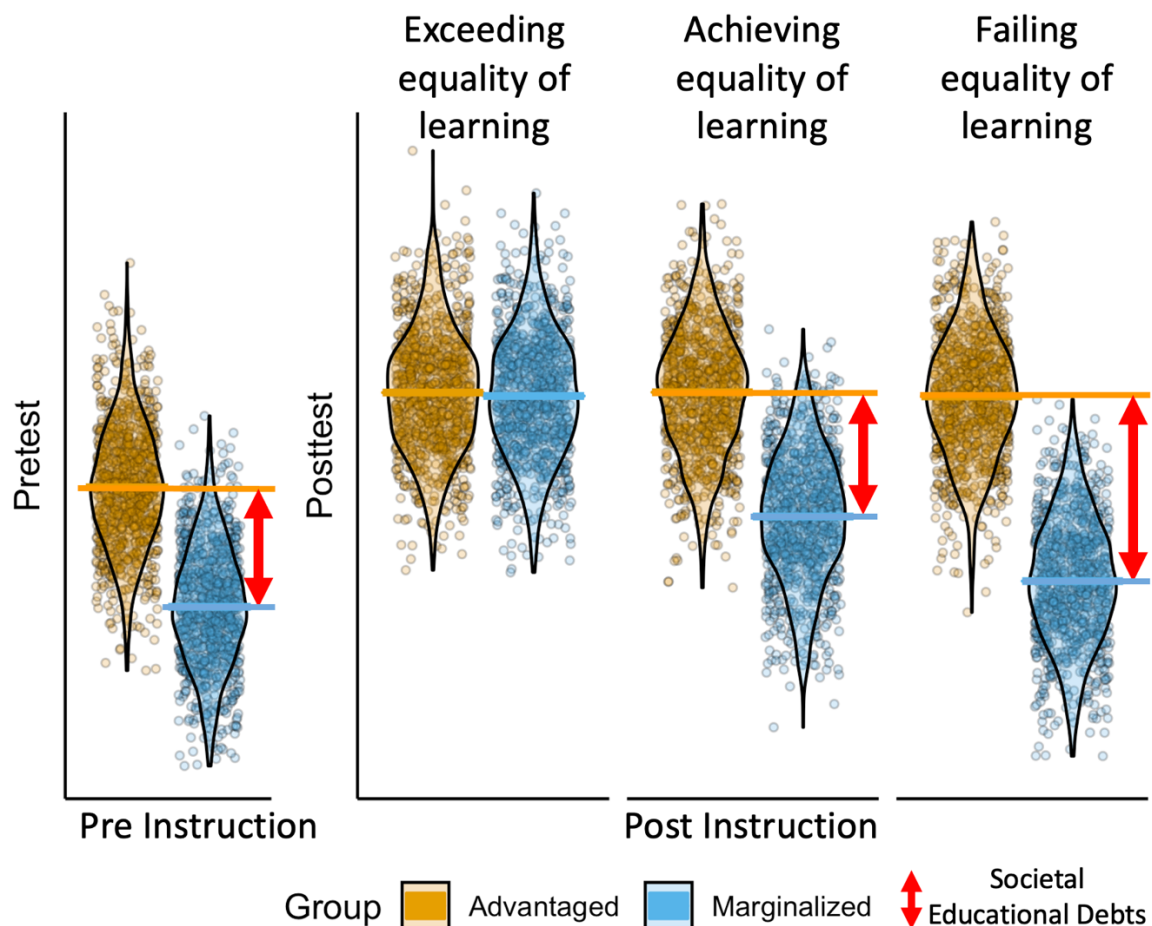

**Figure 1.** A visual representation of society's educational debts before and after instruction with three potential outcomes. The figure uses simulated data with the horizontal lines representing mean scores, dots representing individual students, and the violin plot envelopes representing the density of scores. The figure shows how educational debts can be mitigated (exceeding equality of learning), perpetuated (achieving equality of learning), or exacerbated (failing equality of learning). It also shows that in statistical models, educational debts are measures of average differences between groups, not absolute differences between individuals.

# 4. Model specification

From a QuantCrit perspective, the goals of model specification are to generate models that accurately represent broad scale student outcomes in ways that value their intersectional identities. To accomplish these goals, DBER equity models should maximize two outcomes. 1)

*Accuracy* - minimize bias in predicted outcomes and 2) *precision* – maximize model precision and accurately present uncertainties. In statistics, accuracy and precision are sometimes referred to as bias and variance. Figure 2 shows examples of high and low accuracy and how they relate to over- and under-fitting.

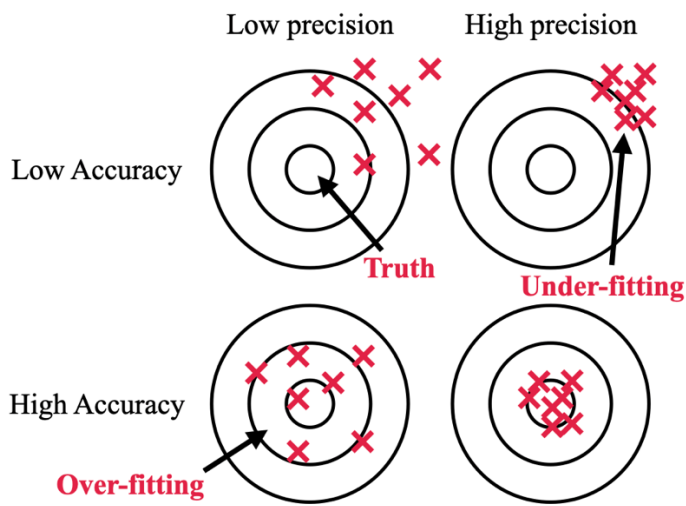

**Figure 2.** A visual representation of accuracy and precision on a dart board. Precision is represented by how close the darts are to each other and accuracy is represented by how close the darts are to the bullseye. Under-fitting models leads to low accuracy, but high precision. Over-fitting models leads to low precision but high accuracy.

Models must balance parsimony and fit (Figure 3; Wagenmakers & Farrell, 2004) to accurately and precisely represent phenomena (Johnson & Omland, 2004). Overly parsimonious models are under-fit and don't include relevant predictor variables, may have biased coefficients, artificially small uncertainties, and seem over-generalizable. Under-parsimonious models are overfit and include non-explanatory predictors, artificially large uncertainties, and limited generalizability (Cawley & Talbot, 2010; Zellner, 2001). Researchers can use large uncertainties for social identifier variables to claim courses have repaid educational debts because the model coefficients for the social identifier variables are no longer statistically significant (for example see, Rodriguez et al., 2012). Statistical power exacerbates the tension between parsimony and fit when developing intersectional models. Statistical power describes a significance test's ability to identify a coefficient with a useful level of precision. The sample size, sub-sample sizes, and number of predictor variables influence a test's statistical power. Intersectional models include interaction terms between variables that can double (e.g., interacting race and gender) or quadruple (e.g., interacting race, gender, and first-generation college status) the number of predictor variables. Dividing the data into more groups decreases the sub-sample sizes. In our recent work, we have used 20 as a minimum sub-sample size for including variables in the model (e.g., at least 20 Hispanic first-generation women; Simmons et al., 2011). This sub-sample size cut-off minimizes the risk of presenting anomalous results with large uncertainties that could be taken out of context to harm minoritized groups.

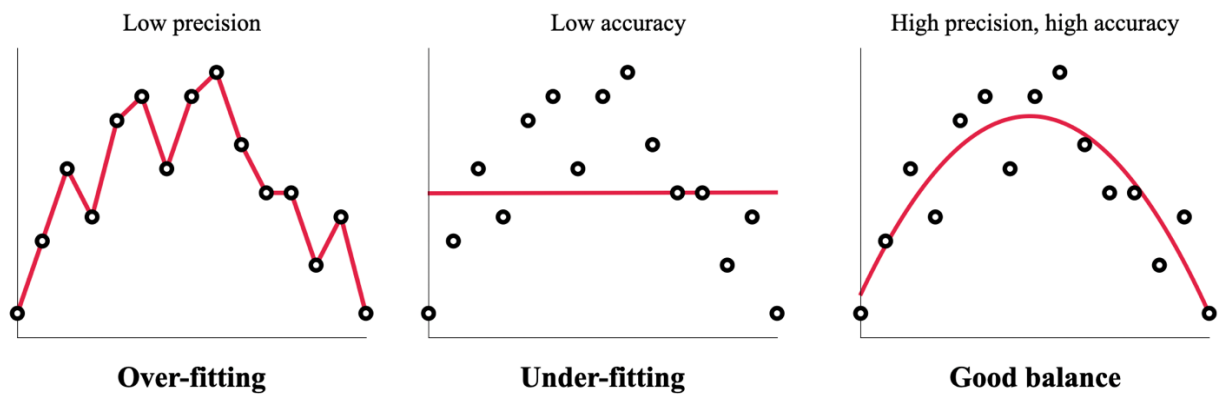

**Figure 3.** Over-fit models capture the noise along with the underlying patter, creating low precision models. Under-fit models fail to capture the underlying pattern, creating low accuracy models. A model with good balance between precision and accuracy captures the underlying pattern in a generalizable manner.

Large datasets are required to power tests with intersectional models. For example, in our review of 60 DBER studies, detailed below, 6 reported their sample sizes disaggregated by gender and race. Black women made up 2% of those samples; a typical study needs to include 1,000 students to include 20 Black women. But only 43% of the reviewed studies had more than 1,000 students.

## 4.1 DBER's use of specification methods

To characterize DBER methods of model specification, we reviewed six DBER journals. The journals cover the disciplines of biology (Cell Biology Education - Life Sciences Education), chemistry (Journal of Chemical Education), engineering (Journal of Engineering Education), math (Journal of Research in Mathematics Education), physics (Physical Review Physics Education Research), and STEM broadly (Journal of Women and Minorities in Science and Engineering). Using each journal's search function, we searched for the term "regression" and "equity." We then identified the ten most recent publications that performed regression analyses of student outcomes. If these two search terms did not identify ten publications, we reran the search only using the term "regression." Seventy-five percent of the articles examined equity, and the date of publications ranged from 1975 to 2020 with a median of 2016.5.

**Table 2.** Model specification methods used and median sample size in 10 publications from each of 6 DBER journals. We will examine each model specification method in detail in the following section.

| Journal | Equity focused | Model specification method | | | | Sample Size Median (min-max) |
|---|---|---|---|---|---|---|
| | | none | p-values | variance explained | AIC | |
| CBE – Life Sciences Education | 9 | 4 | 0 | 1 | 5 | 601 (58-4,810) |
| Journal of Chemical Education | 9 | 9 | 1 | 0 | 0 | 766 (90-3,220) |
| Journal of Engineering Education | 4 | 3 | 5 | 2 | 0 | 5,348 (249-685,429) |
| Physical Review Physics Education Research | 10 | 4 | 4 | 0 | 2 | 1,225.5 (77-47,734) |
| Journal for Research in Mathematics Education | 4 | 8 | 2 | 0 | 0 | 200.5 (16-4,507) |
| Journal of Women and Minorities in Science and Engineering | 10 | 10 | 0 | 0 | 0 | 871 (77-19,073) |
| **Total** | **46** | **38** | **12** | **3** | **7** | **832.5 (16-685,429)** |

Sixty percent of the articles stated their model specification methods. The others required extrapolation from the provided information. Of the 60 articles, 63% did not report using any explicit model specification criteria (Table 2), presumably relying on a priori model specification based on theoretical deductions rather than observations. Twenty percent of the articles used statistical significance, and the least common methods were information criteria (12%) and variance explained (5%).

# 5. Evaluation of specification methods

Each of the four model specification methods can lead to a different best model with unique findings and conclusions. We will examine how well the models from each method aligned with the two goals of model specification discussed in our conceptual framework: 1) *accuracy* - minimize bias in predicted outcomes across diverse student identities and 2) *precision* - accurately reflect the predictions' uncertainty. These examinations look at the general theory for the method and its applications in DBER investigations of equity.

## 5.1 A priori model specification

### 5.1.1 General theory

The most common method for model creation in our literature review was a priori model specification (See Table 2). When a priori identifying a best model, researchers forgo any statistical procedure. This method relies on researchers accurately identifying strong predictors and eliminating poor predictors before running any models. Existing theory and literature can help inform this process (Shmueli, 2010). Even in areas where widely accepted models exist, their use without considering fit criteria has led to poorly specified models (Katsanevakis & Maravelias, 2008).

### 5.1.2 Application in DBER investigations of equity

Articles that we classified as using a priori model fit may have used another method but failed to report their process. These hidden decisions make it difficult to interpret the usefulness and validity of final models produced by studies that don't explicitly state their model specification process. Those that did use a priori methods drew on prior research and theory to decide which variables to include and how to disaggregate social identifier data. DBER equity research, however, often lacks a theoretical framework. Models frequently disaggregate data based on institutional directives, such as the National Science Foundation's definition of underrepresented minorities, and ingrained historical oppression, such as institutional data or scantron sheets that treat gender as a binary (Traxler et al., 2016). Metcalf (2016) argues that these prefabricated variables prioritize statistical significance over meaningful exploration of inequities.

A priori model specification may produce over-specified or underspecified models. Overspecified models may include non-explanatory or redundant predictor variables, which is problematic for the reasons discussed in the Section 4. Model Specification. Underspecified models may not disaggregate social identifier groups or account for intersectionality across social identifier variables. Thus, it is difficult to evaluate if a priori specified models achieve either the first goal (accuracy) of minimizing bias or the second goal (precision) of accurately reflecting model uncertainty.

## 5.2 Statistical significance model specification

### 5.2.1 General theory

Researchers sometimes use statistical significance to identify and remove predictors that do not meet a p-value threshold (typically $p<0.05$). Using a predictor variable's p-value addresses model parsimony by only including the largest coefficients relative to their uncertainty. Researchers implement statistical significance model specification either through an incremental (i.e., stepwise) or single-step process (Kadane & Lazar, 2004). Stepwise model specification adds and removes variables one at a time. The single-step procedure runs the a priori model and then removes all variables that were not statistically significant.

The American Statistical Association, however, has joined a growing group of organizations critiquing the use of statistical significance in the sciences (Amrhein, Greenland, & McShane, 2019; Wasserstein et al., 2019; Wasserstein & Lazar, 2016), including its use in model specification (Raftery, 1995; Shmueli, 2010; Wang, 2019). Researchers calculate predictor variable coefficient p-values from a combination of the coefficient's estimated size (practical significance) and its standard error (uncertainty). This relationship between statistical significance and practical significance has led many to conflate the two (Wasserstein & Lazar, 2016). Authors even incorrectly argue that coefficients with p>0.05 are 'statistically zero' when their studies lack the precision to support such claims. The misunderstanding and misuse of p-values have led some journals (e.g., Basic and Applied Social Psychology) to ban its use in favor of more descriptive measures (e.g., confidence intervals) (Wang, 2019).

## 5.2.2 Application in DBER investigations of equity

Specifying models using statistical significance poses particular problems for DBER investigations of equity. Sample sizes influence which variables achieve statistical significance. Even highly predictive variables will not achieve statistical significance without a large enough sample, and dichotomous social identifier variables require larger samples than continuous variables such as course grades and SAT scores (Astivia et al., 2019). The underrepresentation of minoritized groups in STEM courses and DBER datasets makes it unlikely that practically significant social identifier variables will be statistically significant. Lack of statistical significance for social identifier variables can lead researchers to incorrectly conclude courses or interventions achieved equality between groups. Many DBER equity investigations include predictors that account for differences in students' prior preparation (e.g., pretest score, ACT score, or GPA). These continuous variables further decrease the likelihood of social identifier variables being statistically significant because both are proxies for the same constructs of oppression (e.g., racism, sexism, homophobia, and classism). In other words, regression models that include variables for prior preparation risk overlooking the educational debts society owes minoritized students and risk concluding that unjust course outcomes are equitable. Statistical power limitations in DBER studies, which we discussed earlier, also mean that few studies will have large enough samples to build intersectional models. This can lead researchers to incorrectly conclude that interventions impact all social identifier groups the same. Because of its overly aggressive and biased removal of variables, using statistical significance for model specification is likely to identify an under-specified best model.

Statistical significance model specification leads to under-specified models that exclude informative predictor variables. These models can *fail* the first goal (accuracy) of minimizing bias in predicted outcomes across diverse student identities by indicating differences don't exist when they do. The removal of informative predictor variables creates artificially small standard errors that misrepresent the certainty of the models leading them to *fail* to meet the second goal (precision) of accurately reflecting model uncertainty. For these reasons, we do *not* recommend using statistical significance for model specification in DBER investigations of equity.

## 5.3 Variance explained model specification

### 5.3.1 General theory

Variance explained measures how much the inclusion of predictor variables reduces the variance as a fraction of the total variance and is represented by the $R^2$ term, see Table 1 for further definitions. A model with no predictive ability would explain 0% of the variance. A model that perfectly predicts each datum would have 100% of the variance explained. While the typical variance explained by models varies by field, Cohen (1988) provides a rule of thumb in the social sciences that 2% is the recommended minimum, 13% is moderate, and 26% is a strong effect. Another option for calculating the variance explained is adjusted $R^2$, which includes an adjustment for the number of variables. Researchers can also calculate the additional variance explained for each variable, reflecting the amount the variance explained improved with the variable's inclusion. However, a shortcoming of additional variance explained is its dependency on the order that researchers include the variables in the model.

Researchers typically use variance explained for model specification in two ways: 1) maximizing variance explained or 2) using an additional variance explained cut-off. The first way identifies the model with the most variance explained as the best model (Johnson & Omland, 2004). This method maximizes model fit and ignores parsimony because even the inclusion of a variable unrelated to the outcome measure rarely decreases the variance explained. The lack of a mechanism for creating model parsimony usually leads to identifying the a priori model as the best model. The second method includes an additional variance explained cut-off for variables. In this method, variables are introduced one at a time (i.e., stepwise) and only kept if they explain additional variance above the cut-off value. These cut-offs vary by study. In the study that we re-analyzed, Van Dusen & Nissen (2020) required that each variable improve the combined level-1 and level-2 percentage variance explained by 1%. Unlike the first method, the cut-offs in this method support model parsimony.

### 5.3.2 Application in DBER investigations of equity

Specifying models by maximizing variance explained using $R^2$ typically leads to the same best model as a priori model specification and the same drawbacks as that method. The second method for using variance explained in model specification (additional variance explained) and using adjusted $R^2$ are functionally equivalent.

Specifying models using the second method of variance explained (additional variance explained) has two primary issues. First, the order that researchers add variables can impact their additional variance explained. For example, predictor variables for prior preparation and social identifiers will explain some of the same variance in an outcome shaped by systemic oppression. However, researchers and policymakers can focus on the prior preparation to ignore the oppression and inequalities the data reveals. Depending on the procedures and software used, whichever predictor variable is added to the model first will account for their shared variance, thereby boosting its additional variance explained and decreasing the additional variance explained for the variables added later. For example, if DBER investigations of equity add prior performance variables first, then they are less likely to include social identifier variables in the best model. This dependency on the order of variable addition allows researchers to influence the process and circumvent the purpose of model specification.

The second issue is that the relationship between the variance explained by a variable and the variables size are complex. Variance explained depends on complex interactions between the sample sizes, coefficient sizes, and data distributions (Ferguson, 2009). Imbalanced samples, such as for equity models with small subsamples from minoritized groups, may require that social identifier predictor variables have large coefficients representing extreme inequalities to meet cutoffs set by researchers. To mitigate these issues, researchers can create proportional samples from the larger dataset. This data manipulation, however, negatively impacts statistical power.

Variance explained tends to under-specify models by not including informative predictor variables. These models *fail* the first goal (accuracy) of minimizing bias in outcomes across diverse student identities by representing outcomes as homogenous. The lack of relevant predictor variables is also likely to create artificially small standard errors, leading them to *fail* to meet the second goal (precision) of accurately reflecting model uncertainty. For these reasons, we do *not* recommend using variance explained for model specification in DBER investigations of equity.

# 5.4 Information criterion model specification

## 5.4.1 General theory

Information criteria are grounded in information theory and take both model fit and parsimony into account (Zucchini, 2000; Raftery, 1999). One of the more popular information criteria is the Akaike information criterion (AIC; Akaike, 1998; Wagenmakers & Farrell, 2004). AIC estimates the Kullback–Leibler information lost by approximating full reality with the fitted model to establish a goodness of fit (Johnson & Omland, 2004). AIC accomplishes this by calculating the lack of fit with a penalty for additional predictors. AIC corrected (AICc), which we used, includes a correction term for small sample sizes. Other information criteria include Bayesian information criterion, Watanabe–Akaike information criterion, and leave-one-out cross-validation. Each information criterion calculates model fit and parsimony in slightly different ways but identifies best models in similar ways. AIC puts a stronger emphasis on model fit than other information criteria (e.g., Bayesian Information Criterion) by including a smaller penalty for additional predictors.

To identify a best model, researchers compare the information criterion for each proposed model. AIC values can be quite large. Nonetheless, differences in the single digits inform the quality of model fit. Burnham and Anderson (2002) advise that models within 2 points of the lowest AICc value have equally strong fits and models within 8 AICc points are still worth considering. This gradient of support, rather than a single cut-off value, gives researchers some flexibility in identifying an empirically and conceptually supported best model. The lack of dependence on the order that researchers add variables allows for the automation of model specification. For example, the dredge function (Barton & Barton, 2015) calculates information criterion for every possible combination of predictor variables in a model.

### 5.4.2 Application in DBER investigations of equity

Because our conceptual framework forefronts intersectional models, we recommend using an information criterion with a smaller penalty for including additional variables (e.g., AICc). AICc runs the risk of overspecifying models, but alternate information criteria can be used to adjust the emphasis on model parsimony. Information criterion's robust model specification methods allow it to handle the range of sample sizes and unbalanced samples common in DBER investigations of equity in ways that minimize model bias and uncertainty.

Information criterion will identify models that include predictor variables that provide unique information. These models *meet* the first goal (accuracy) of minimizing bias in predicted outcomes across diverse student identities by including informative variables. These models' inclusion of informative and exclusion of non-informative predictor variables leads them to *meet* the second goal (precision) of accurately reflecting model uncertainty. For these reasons, we recommend using information criterion for model specification in DBER investigations of equity.

## 5.5 Summary of model specification methods

Based on the review of the literature, information criterion provide a method for model specification that aligns with QuantCrit because it allows for model parsimony without removing social identifier variables due to small sample sizes. A priori models do not allow for model parsimony. P-value and variance explained model specification both tend to remove predictor variables for social identifier groups with small sample sizes. This bias against small samples can undermine research seeking to understand the effects of intersecting axes of oppression, which require disaggregating the data across multiple underrepresented social identity groups.

# 6. Worked example

In our worked example, we reanalyze data from Van Dusen & Nissen (2020) using the four model specification methods and comparing the findings from each. The purpose of this worked example is to illustrate the ways that each method can impact model specification, the findings that come from them, and the implications for equity. In this example, we compare the other models against the information criterion model as the methods used to create it are best supported by the literature and our QuantCrit perspective.

As we followed the original publication's procedures for data collection and processing, handling missing data, and creating descriptive statistics, those subsections are truncated versions of the original article. While we agree with most of the methods used in that study, our ongoing use of QuantCrit has led us to identify several areas where we would handle gender and race data differently than we did in our 2020 article, which we discuss in the Section 6.3.4

Reflections on handling social identifier data. To maintain fidelity to the study, however, we follow the methods used in the original investigation.

## 6.2 Protection of vulnerable populations

We analyzed an existing dataset from the Learning About STEM Student Outcomes (LASSO) database under California State University Chico IRB protocol #5582. LASSO is a low-stakes online assessment platform that is free for instructors to use. The data only included students who consented to share their de-identified data with researchers. LASSO removed all student, course, instructor, and institution identifiers from the data we analyzed. To further ensure vulnerable populations' safety, McKensie Mack of Radical Copy performed an equity audit of the manuscript.

## 6.3 Methods

### 6.3.1 Data collection & processing

Data came from the Learning About STEM Student Outcomes (LASSO) platform's anonymized research database (Van Dusen, 2018). Instructors used the LASSO platform to administer and score the research-based assessments (RBAs) online. The RBAs were the Force Concept Inventory (FCI; Hestenes & Swackhamer, 1992) and the Force and Motion Conceptual Evaluation (FMCE; Thornton & Sokoloff, 1998). The FCI and FMCE use multiple choice questions to assess core concepts around forces and motion in first-semester physics courses. Both have absolute gains from pretest to posttest between 10% and 30% (Rodriguez, 2012). The data included 15,267 students in 201 courses from 31 institutions. The dataset had students' self-identified gender, race, ethnicity, and if they were retaking the course; student pretest score, posttest score, time spent taking the assessment; and whether the course used collaborative learning.

We cleaned the data by removing students if they were missing data for their gender or retaking the course. We then removed individual tests if the student took less than 5 minutes on the assessment or completed less than 80% of the questions. If a student had neither a pretest nor a posttest score after these filters, we removed their data. We removed courses with less than 40% student participation on either the pretest or posttest from the data and courses with fewer than either eight pretests or eight posttests. After filtering, 58% of the students had matched pretest and posttest scores, which fell in the range of typical participation rates in the literature (Nissen et al., 2018).

After cleaning the data, we performed hierarchical multiple imputations (HMI) with the hmi (Speidel et al., 2018) and mice (van Buuren & Groothuis-Oudshoorn, 2010) packages in R-Studio V. 1.1.456 to address missing data. The multiply imputed dataset used in this study was the same one used in the earlier study.

We analyzed the data using hierarchical linear models (HLM). All models were 2-level hierarchical linear models that nested students within classes. Each of the models used fixed slopes and random intercepts. The models were regressed using the maximum likelihood estimates in the lme4 package (Bates et al., 2010) in R.

### 6.3.2 Gender, race, and ethnicity

We used students' self-reported social identifier data collected through the LASSO platform to categorize their gender, race, and ethnicity. We entered each social identifier variable into the model using a placeholder (0/1) variable. This section will first discuss how we handled the social identifier data in the prior study and this study. We will then discuss how we have modified these practices in later investigations.

### 6.3.3 Social identifiers in this and the prior study

As discussed earlier, we removed students from the dataset who did not reply to the gender question. We aggregated the students who selected transgender, other, or typed in a gender with women. Aggregating the data in this way reduced students' gender data to either men or women. Race and ethnicity categories included Black, Hispanic, Asian, Hawaiian or Pacific Islander (Pac. Islander), other, or White based on student responses to the social identifiers questionnaire. The other category included students who selected other, Alaskan Native or American Indian, or did not respond. We included the Alaskan Native and American Indian students in the other category because of sample size (N=61). We identified students with multiple racial identities by the identity with the smallest sample size to make the race/ethnicity categories independent of one another to simplify the model and preserve statistical power. For example, a student who identified as both Black and Hispanic was included as a Black student because that was the smaller sample size. Except for Hispanic students, the number of students of multiple races/ethnicities was small.

### 6.3.4 Reflections on handling social identifier data

Van Dusen & Nissen's (2020) decisions preserved statistical power for using variance explained for model selection. In later work (e.g., Nissen, Her Many Horses, & Van Dusen, 2021; Van Dusen et al., 2021), we focused on respecting students' identities first and following statistical guidance for minimum group sizes second. We used a minimum group size of 20 (Simmons, Johnson, & Simonsohn, 2011) to include social identifier variables or interactions to minimize the chance of spurious results. For gender, we included variables for man, woman, transgender, and nonbinary if the number of students who identified with that group reached 20. For each student, we included all identities that met this threshold. For example, if the data included ten transgender men and ten transgender women, we would include a variable for transgender. Still, we would not include an interaction term (i.e., transgender*woman) until the number of transgender men and women was each 20 or more. For groups with less than 20, we aggregate their data into the variable *other gender*. Aggregating groups with small sample sizes allowed us to keep them in our analyses, which accurately represented the courses they were in while also acknowledging that quantitative measures cannot provide accurate and precise information about small groups.

We have also made several changes in how we model race. Like how we included gender in the model, we used a minimum size of 20 to include a variable or interaction in the model. For groups with less than 20 members in the dataset, we included their data in the other race category for the same reason stated for gender. We stopped restricting each student to one racial identity. When the number of students with two identities reached or exceeded 20, we included an interaction term in the a priori model, e.g., White Hispanic students.

The capabilities and limitations of quantitative models put them in tension with critical theory's tenet of intersectionality. Methodological choices can both cause and mitigate harm with no clear guidance to fully address this tension. However, this tension motivates us to refine and better align our methods with the tenets of critical theory and those tenets provide guidance for making decisions when facing uncertainty.

## 6.3.5 Descriptive statistics

The data set included 187 courses: 153 courses used collaborative instruction, had 11,740 students, and mean gains of 20.9%, and 34 courses used lecture-based instruction, had 2,117 students and mean gains of 15.7%. Forty-eight of the courses used the FMCE. One hundred thirty-nine of the courses used the FCI. We calculated descriptive statistics for the dataset to characterize differences between the mean pretest scores, posttest scores, and gains across student social identifiers (Table 3). The final sample included students and courses from 31 institutions: 20 granted doctorates, six granted master's degrees, three granted bachelor's degrees, and two granted associate degrees.

**Table 3.** Descriptive statistics for the sample, disaggregated by race and gender[4].

| Race/ Ethnicity | Instruction | Gender | N | Gain | |
|---|---|---|---|---|---|
| | | | | Mean | SD |
| Asian | Collab. | Men | 811 | 19.2 | 20.6 |
| | | Women | 611 | 18.6 | 19.8 |
| | Lecture | Men | 96 | 20.1 | 21 |
| | | Women | 75 | 14.9 | 17.7 |
| Black | Collab. | Men | 169 | 16.2 | 19.1 |
| | | Women | 180 | 17.8 | 19.9 |
| | Lecture | Men | 28 | 20.3 | 16.9 |
| | | Women | 36 | 16.4 | 15 |
| Hispanic | Collab. | Men | 1192 | 19.1 | 19.3 |
| | | Women | 568 | 17.4 | 19 |
| | Lecture | Men | 204 | 13.9 | 18.7 |
| | | Women | 1982 | 10.8 | 14.7 |
| Pacific Islander | Collab. | Men | 69 | 19.8 | 17.9 |
| | | Women | 42 | 23.3 | 19.5 |
| | Lecture | Men | 5 | 18.1 | 28.1 |
| | | Women | 10 | 13.5 | 12 |
| Other | Collab. | Men | 497 | 18.1 | 19.2 |
| | | Women | 357 | 19.1 | 19.7 |
| | Lecture | Men | 92 | 14.4 | 21.4 |
| | | Women | 95 | 11.1 | 13.8 |
| White | Collab. | Men | 4791 | 22.1 | 19.2 |
| | | Women | 2429 | 22.9 | 19.5 |
| | Lecture | Men | 593 | 16.7 | 18.9 |
| | | Women | 685 | 16.8 | 18.2 |

## 6.3.6 Model development

To develop our gain (posttest - pretest) models, we began by creating a model with no predictor variables (i.e., an unconditional model) and then used a forward stepwise process to add or remove predictor variables incrementally. We created 11 total models. Table 4 shows the full list of variables. To ease interpretation, we group mean-centered (i.e., centered on the mean prescore in the course) student prescore variable, grand mean-centered (i.e., centered on the mean across all courses) the class mean prescore, and left the rest of the variables uncentered.

[4] Other races and genders are included in the dataset and model, but their sample sizes were not large enough to ethically include them in this table.

**Table 4.** Predictor variables examined, their level, and whether they were centered.

| Variable | Level | Centering |
|---|---|---|
| Student prescore | Student | Group mean-centered |
| Class mean prescore | Course | Grand mean-centered |
| Retake | Course | none |
| Collaborative | Course | none |
| Woman | Student | none |
| Asian | Student | none |
| Black | Student | none |
| Hispanic | Student | none |
| Race_other | Student | none |
| Pacific Islander | Student | none |

### 6.3.7 Model selection

Each model specification method identified a different best model. Using the statistical significance procedure, we identified the model with all statistically significant coefficients as the best model. To specify the model using variance explained, we calculated the percentage of the level-1 and level-2 variance explained by each model. We then examined whether the percentage of level-1 variance explained plus the percentage of level-2 variance explained improved by at least 1%. If so, we deemed the new model better than the preceding model. This process continued until new variables failed to improve the combined variable explained by at least 1%. To use the information criteria, we calculated the AICc score for each model. We then identified the model with the lowest AICc score and calculated the difference in AICc scores between that model and each other model. The a priori model included all of the variables considered in Van Dusen & Nissen (2020).

## 6.4 Findings

We began the model development process by adding the student background variables to isolate their effects before adding our variables of interest (i.e., social identifiers and course type; see Table 5). Model 1 was the unconditional model, which predicted gains without using any predictor variables. Model 2 added a predictor variable for student pretest scores. Model 3 added a predictor variable for the course's mean pretest score. As none of the selection criteria indicated that the class mean prescore was a useful predictor, we dropped Model 3. We built Model 4 off Model 2 by adding a variable for whether a student was retaking the course. Model 5 added a predictor variable for whether the course was taught using collaborative learning activities. Model 6 added a variable for gender. Model 7 added five variables for race (Asian, Black, Hispanic, Hawaiian or Pacific Islander, and other). We built Model 8 specifically for the statistical significance criteria by removing the non-statistically significant predictor (collaborative) from Model 7. Model 9 built off of Model 7 by adding the interaction between gender and race. Model 10 added interaction terms for collaborative, gender, and race. Model 11 included all the variables that were a priori identified as being potentially informative.

**Table 5.** A table of the model specification process with values for delta $R^2$ and delta AICc, whether all variables were $p<0.05$, and which model was identified as best by each model specification procedure. We use bold to indicate the combined percentage variance explained changes that showed improvement in each model. The Delta AICc scores are referenced against the model with the lowest AICc score (Model 10).

| Model | Coefficients | $p$-value criteria | Delta $R^2$ | Delta AICc | Best Model |
|---|---|---|---|---|---|
| 1 | Intercept | met | 0.00% | 1544.1 | - |
| 2 | Int. + Pre_student | met | **7.74%** | 237.2 | - |
| 3 | Int. + Pre_s. + Pre_class | - | -0.49% | 242.8 | - |
| 4 | Int. + Pre_s. + retake | met | **4.22%** | 201.8 | - |
| 5 | Int. + Pre_s. + retake + collaborative | - | **2.06%** | 197 | - |
| 6 | Int. + Pre_s. + retake + collaborative + gender | - | **1.02%** | 167.5 | - |
| 7 | Int. + Pre_s. + retake + collaborative + gender + race | - | **2.79%** | 41.4 | $R^2$ |
| 8 | Int. + Pre_s. + retake + gender + race | met | -1.59% | 46.2 | p-value |
| 9 | Int. + Pre_s. + retake + collaborative + gender*race | - | -0.14% | 27.6 | - |
| 10 | Int. + Pre_s. + retake + collaborative*gender*race | - | -0.26% | min | AICc |
| 11 | Int. + Pre_s. + retake + collaborative*race*gender + pre_class | - | -0.14% | 4.7 | a priori |

Table 6 shows the variables used in each of the best models with their coefficients, p-values, and standard error. The table also shows the variance at both level-1 and level-2 and the AICc scores.

**Table 6.** The four models identified as the best by the three specifications criteria and the a priori model.

| Variable | Statistical Significance Best Model (Model 8) | | | Variance Explained Best Model (Model 7) | | | Information Criteria Best Model (Model 10) | | | A Prior Best Model (Model 11) | | |
|---|---|---|---|---|---|---|---|---|---|---|---|---|
| | Coeff. | S.E. | p-value | Coeff. | S.E. | p-value | Coeff. | S.E. | p-value | Coeff. | S.E. | p-value |
| Intercept | 19.7 | 0.817 | 0.000 | 17.2 | 1.55 | 0.000 | 16.9 | 1.65 | 0.000 | 16.7 | 1.67 | 0.000 |
| Student prescore | -0.328 | 0.011 | 0.000 | -0.328 | 0.011 | 0.000 | -0.329 | 0.011 | 0.000 | -0.329 | 0.011 | 0.000 |
| Retake | 2.77 | 0.603 | 0.000 | 2.74 | 0.602 | 0.000 | 2.72 | 0.606 | 0.000 | 2.72 | 0.606 | 0.000 |
| Woman | -1.87 | 0.438 | 0.000 | -1.87 | 0.437 | 0.000 | -1.37 | 1.17 | 0.245 | -1.39 | 1.17 | 0.237 |
| Asian | -1.92 | 0.51 | 0.000 | -1.93 | 0.51 | 0.000 | -2.62 | 2.11 | 0.213 | -2.62 | 2.11 | 0.214 |
| Black | -5.88 | 1.08 | 0.000 | -5.89 | 1.08 | 0.000 | -6.16 | 4.26 | 0.153 | -6.15 | 4.26 | 0.153 |
| Hispanic | -4.21 | 0.527 | 0.000 | -4.21 | 0.527 | 0.000 | -3.27 | 1.57 | 0.038 | -3.3 | 1.57 | 0.036 |
| Race_other | -3.1 | 0.65 | 0.000 | -3.08 | 0.65 | 0.000 | -2.6 | 2.17 | 0.232 | -2.62 | 2.17 | 0.229 |
| Pac. Islander | -4.4 | 1.88 | 0.021 | -4.41 | 1.88 | 0.021 | -5.02 | 9.8 | 0.611 | -5.07 | 9.8 | 0.607 |
| Collaborative | - | - | - | 3.1 | 1.61 | 0.054 | 3.61 | 1.74 | 0.038 | 3.87 | 1.76 | 0.028 |
| Woman*Asian | - | - | - | - | - | - | -0.347 | 3.09 | 0.91 | -0.35 | 3.09 | 0.91 |
| Woman*Black | - | - | - | - | - | - | 4.62 | 5.24 | 0.38 | 4.61 | 5.24 | 0.381 |
| Woman*Hispanic | - | - | - | - | - | - | -0.67 | 2.28 | 0.769 | -0.656 | 2.28 | 0.774 |
| Woman*Race_other | - | - | - | - | - | - | -0.993 | 3.31 | 0.765 | -0.984 | 3.31 | 0.767 |
| Woman*Pac. Islander | - | - | - | - | - | - | -0.065 | 11.9 | 0.996 | -0.028 | 11.9 | 0.998 |
| Coll.*Woman | - | - | - | - | - | - | -0.941 | 1.23 | 0.443 | -0.929 | 1.23 | 0.45 |
| Coll.*Asian | - | - | - | - | - | - | 0.201 | 2.24 | 0.928 | 0.179 | 2.24 | 0.938 |
| Coll.*Black | - | - | - | - | - | - | -1.5 | 4.21 | 0.723 | -1.55 | 4.21 | 0.714 |
| Coll.*Hispanic | - | - | - | - | - | - | -0.74 | 1.72 | 0.666 | -0.732 | 1.72 | 0.67 |
| Coll.*Race_other | - | - | - | - | - | - | -1.64 | 2.49 | 0.513 | -1.63 | 2.49 | 0.514 |
| Coll.*Pac. Islander | - | - | - | - | - | - | -0.478 | 10.2 | 0.963 | -0.441 | 10.2 | 0.966 |
| Coll.*Woman*Asian | - | - | - | - | - | - | 1.77 | 3.27 | 0.589 | 1.78 | 3.28 | 0.587 |
| Coll.*Woman*Black | - | - | - | - | - | - | -1.94 | 5.54 | 0.727 | -1.92 | 5.53 | 0.729 |
| Coll.*Woman*Hispanic | - | - | - | - | - | - | -0.377 | 2.58 | 0.884 | -0.383 | 2.58 | 0.882 |
| Coll.*Woman*Race_other | - | - | - | - | - | - | 3.86 | 3.75 | 0.308 | 3.86 | 3.76 | 0.308 |
| Coll.*Woman*Pac. Islander | - | - | - | - | - | - | 3.09 | 12.6 | 0.807 | 3.05 | 12.6 | 0.809 |
| Class mean pre | - | - | - | - | - | - | - | - | - | -0.061 | 0.065 | 0.350 |
| $R^2$ - course | | 5.47% | | | 7.46% | | | 7.03% | | | 6.88% | |
| $R^2$ - student | | 10.37% | | | 10.37% | | | 10.40% | | | 10.40% | |
| AICc | | 118418.6 | | | 118413.8 | | | 118372.4 | | | 118377.1 | |

In the remainder of the findings section, we examined what each of the four best models predicted about student outcomes and what conclusions about classroom inequities they supported. To focus the findings on how the model specification impacted conclusions about equity, we've only plotted outcomes for the four most populous racial groups in our data (Black, Hispanic, Asian, and White). The small number of students who identified as Hawaiian or Pacific Islander in lecture-based courses meant that some of the models had large uncertainty about the outcomes for that group of students.

To compare findings across model specification methods (Figure 4), we discussed each model's findings in five areas: 1) collaborative learning, 2) sexism, 3) racism, 4) intersectionality (i.e., gender*race), 5) equity and collaborative learning (i.e., collaborative*social identifiers), and 6) shortcomings. Our discussion of the four models is informed by our conceptual framework (e.g., we identify racism and sexism as the causes of inequitable outcomes). We discussed the models in order of least to most complex. Additional examination of the model shortcomings and implications for their use are included in the discussion section.

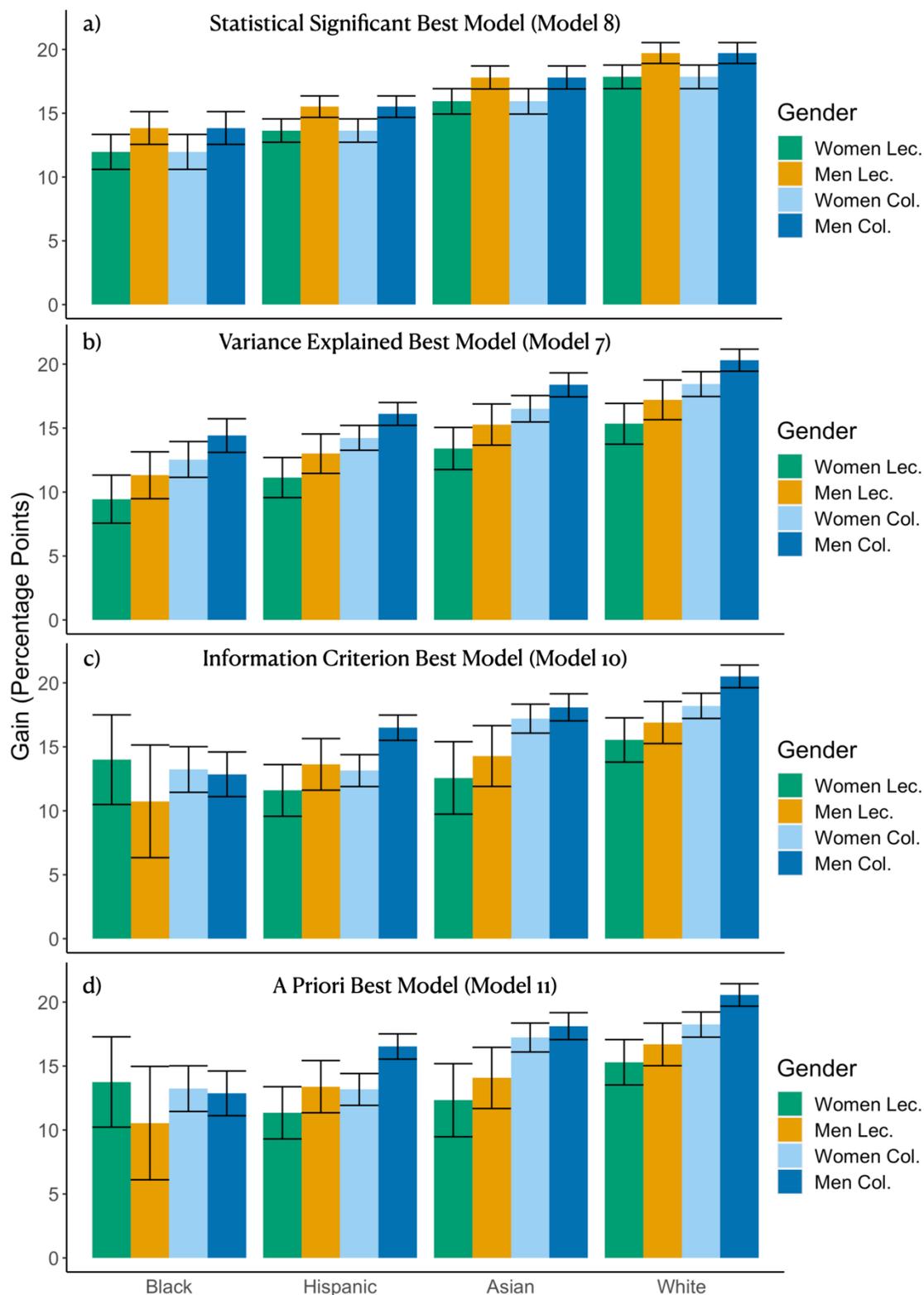

**Figure 4.** Predicted gains across social identifier groups for the four best models: a) statistical significance, Model 7; b) variance explained, Model 8; c) information criterion, Model 10; and d) a priori, Model 11. Error bars represent +/- 1 standard error.

### 6.4.1 Statistical significance best model

The statistical significance model selection criteria identified Model 8 as the best model. Model 8 included social identifier variables for gender and race but no interaction effects or collaborative instruction variables. Model 8 indicated:

1. Collaborative learning - Conceptual learning in courses with collaborative learning and lecture-based instruction were identical.
2. Sexism - The gender variable indicated sexism negatively impacted all women's learning gains by 1.9 percentage points, an 11% lower learning gain than would occur in an equitable course.
3. Racism - Racism negatively impacted Asian, Black, and Hispanic students; racism's negative impacts were largest for Black students and the smallest for Asian students.
4. Intersectionality - There were no intersectional effects. The impacts of racism were constant across genders, and sexism's impacts were consistent across racial groups.
5. Equity and collaborative learning - Collaborative learning was not associated with any shifts in equality of learning.
6. Shortcomings – The model's lack of a term for collaborative learning makes it impossible to identify whether collaborative learning is associated with improved average student learning gains. The model's lack of interaction terms between collaborative learning and social identifier variables makes it impossible to identify whether collaborative learning is associated with increased or decreased equality of learning. The model's lack of interaction terms between gender and race variables makes it impossible to identify any intersectional effects. The general lack of interaction terms in the model also produced artificially small error bars for groups that were not well represented in the dataset (e.g., Black women and men in lecture-based courses).

### 6.4.2 Variance explained best model

The variance explained model selection criteria identified Model 7 as the best model. Model 7 was the same as Model 8 with one exception; it included a variable for collaborative learning. Model 7 indicated:

1. Collaborative learning - Students learned 3.1 percentage points more, a 23% increase in learning gains, in collaborative-based courses.
2. Sexism - Sexism negatively impacted women's learning gains by 1.9 percentage points, a 12% lower learning gain than men.
3. Racism - Racism negatively impacted Asian, Black, and Hispanic students; racism's negative impacts were largest for Black students and the smallest for Asian students.
4. Intersectionality - There were no intersectional effects. The impacts of racism were constant across the genders, and sexism's impacts were consistent across racial groups.
5. Equity and collaborative learning - Equality of learning did not broadly occur in courses that used collaborative learning.
6. Shortcomings – The model's lack of interaction terms between collaborative learning and social identifier variables makes it impossible to identify whether collaborative learning is associated with increased or decreased equality of learning. The model's lack of interaction terms between gender and race variables makes it impossible to identify any intersectional effects. The general lack of interaction terms in the model also produced artificially small error bars for groups that were not well represented in the dataset (e.g., Black women and men in lecture-based courses).

### 6.4.3 Information criterion best model

The information criterion identified Model 10 as the best model. Model 10 included interactions between gender, race, and collaborative learning. Model 10 indicated:

1. Collaborative learning - Overall, students learned more in collaborative-based courses, but this instructional strategy's advantages were not certain for all students. Students learned at least 2.6 percentage points more, a 17% or more increase in learning gains, in collaborative-based courses for all social identifier groups other than Black men, Black women, and Hispanic Women. For Black students, uncertainty in the learning gains in lecture-based courses made comparisons problematic. For Hispanic women, the difference was only 1.5% and was well within the model uncertainty.

2. Sexism - Sexism negatively impacted White and Hispanic women's learning gains by more than 1.3 percentage points in lecture-based courses and more than 2.3 percentage points in collaborative-based courses. Sexism likely harmed Asian women's gains, but more evidence was needed to support strong claims about this harm due to the model's uncertainty. Black men and women had similar learning gains in collaborative courses. The model's uncertainty means that it did not rule out sexism harming Black women's learning gains, but the model did indicate those harms at worst would have been similar to the effects of sexism on Asian, Hispanic, and White women. The uncertainty was too large for Black students in lecture-based courses to make claims about the impacts of sexism on Black women in that course type. The negative impact of sexism in collaborative courses had less uncertainty than in traditional courses due to the prevalence of collaborative courses in the data.

3. Racism - Racism negatively impacted Asian, Black, and Hispanic students to varying amounts. In lecture-based courses, Asian, Black, and Hispanic students had similar negative impacts from racism on their gains (~3.4 percentage points). In collaborative courses, the adverse effects of racism on gains were largest for Black students (6.3 percentage points) and smallest for Asian students (1.7 percentage points).

4. Intersectionality - The model indicated the intersectional effects of racism and sexism were less than the additive effects of both for each racial and gender group. The mediational effect of sexism on racism in this context could be seen in the smaller spread of scores due to racism for women than men in both traditional (3.9% vs. 6.2%) and collaborative learning courses (5.1% vs. 7.6%).

5. Equity and collaborative learning - The impact of collaborative learning varied by social identifier group, but there was no overall evidence of improvement in equality of learning.

6. Shortcomings – The larger error bars in comparison to the prior models stands out as a drawback to this model because it makes it more difficult to identify robust differences. The larger error bars, however, more accurately represent the models' certainty about findings given the variation across sub-group sample sizes. For example, Models 7 and 8 in Figure 4 give similar uncertainties for Black students in both collaborative and lecture based instruction though the much smaller number of students in lecture based courses, approximately one fifth, indicates much greater uncertainty about outcomes for Black women and men in lecture-based courses than in collaborative courses. The error bars in Model 10 provide us a guideline for how confidently the model can supporting different claims based on the sample sizes.

### 6.4.4 A priori best model

The a priori model selection identified Model 11 as the best model. Model 11 was the same as Model 10 with one exception; it included a variable for the class's mean pretest score. The differences between Model 10 and Model 11's predicted outcomes were small for both the predicted gains (<0.25 percentage points) and standard errors (<0.04 percentage points). We did not list out the findings for the first five areas for the a priori best model for brevity's sake as they would be identical to those of the information criterion's best model. The shortcomings of this model specification process were not apparent in our example. Specifically, including additional variables can lead to larger error bars but it did not occur in this case.

# 7. Discussion

Our worked example had a sample size (15,267 students) ~18 times larger than the median sample size from our literature review of DBER investigations of equity (833 students), and only 6 studies we reviewed reported larger samples. Our worked example found the best models identified by statistical significance (Model 8) and variance explained (Model 7) both performed as predicted, leaving out relevant predictor variables and creating underspecified models. These models failed to meet the accuracy and precision goals we identified for DBER investigations of equity. We can find evidence for their failure to meet the accuracy goal (minimizing bias in predicted outcomes across diverse student identities) in the disconnect between the descriptive statistics and the model predictions. For example, the descriptive statistics found White men to have gains of 16.7% in lecture-based courses, while the statistical significance model predicted their gains at 19.7%. This contrasts with the information criterion model which predicted their gains to be 16.9%. Evidence for their failure to meet the precision goal (accurately reflecting model uncertainty) can be found in the disconnect between sample distribution and the standard errors. As standard error is inversely proportional to the sample size's square root, the standard error should be larger for social identifier groups with smaller sample sizes. However, in these models, White men's standard error is nearly identical to that of Black women, despite having ~25 times more White men in the sample (5,384 vs. 216).

The worked example also illustrated the unreasonableness of using p-values as a rigid go/no-go test. The variable for collaborative learning had a p-value of 0.054 in Model 7, leading to its removal in the statistical significance best model (Model 8). The variance explained best model (Model 7), however, showed that collaborative learning was associated with a meaningful average increase of 3.1 percentage points in student gains (an average 23% increase in gains). While it is common for statistical significance to be used in such ways, any strict rules for model specification that do not take a more holistic approach run the risk of hiding meaningful relationships and differences.

In this example, the a priori best model (Model 11) and information criterion best model (Model 10) were nearly identical. As it happened, almost all of the variables identified a priori as being potentially relevant were, leading to little differentiation between these two models. Both were intersectional and met the modeling goals we identified for DBER investigations of equity. The fact that both models led to such similar results is no surprise given that Model 11's AICc score was only 4.7 points higher than Model 10s, identifying it as a reasonable model to use. The lack of predictor variables such as test scores, course grades and college or high school GPA among others limits this study's ability to differentiate between the information criterion and a prior model selection processes. While both of these methods led to similar findings in this case, the a priori method acted as we predicted and identified a variable as relevant that none of the other methods identified as relevant. Not all investigations will have a substantial similarity between the a priori and information criterion best models. A priori identifying the best model runs the risk of creating over-specified models with misleading conclusions and over large error terms.

# 8. Conclusions

Critical theory tells us that presenting statistical analyses and their findings as objective and fact-based obscures the biases that data collection and analyses methods introduce. These hidden biases are particularly damaging in quantitative investigations of classroom equity because the practices are rooted in a tradition of hegemonic oppression (Zuberi & Bonilla-Silva, 2008). We performed this investigation of model specification methods to support future DBER investigations of equity in acknowledging and mitigating these biases. Our QuantCrit perspective led us to combine statistics emphasis on model fit and parsimony with critical theory's emphasis on intersectionality to create two goals for model specification in DBER investigations of equity: 1) accuracy - minimize bias in outcomes across diverse student identities and 2) precision - accurately reflect the prediction's uncertainty.

Our examination of DBER's four standard model specification methods led us to conclude that a priori best models lack of parsimony criteria is likely to create over-specified models, unnecessarily increasing model uncertainty and failing the model precision goal. Model specification using statistical significance or variance explained can over-emphasize model parsimony and create under-specified models that eliminate intersectionality from the model and fail both the model accuracy and precision goals. Information criterion strikes a balance between model fit and parsimony, allowing models to be intersectional while minimizing model uncertainty, thereby meeting the model accuracy and precision goals. We conclude that using information criterion provides the best method for specifying models in DBER investigation of equity. While exploring the different information criteria was beyond this research's scope, we recommend using AICc as the information criterion metric. AICc is commonly used and puts less emphasis on parsimony than most other information criteria making them more likely to include intersectional effects in their best models.

Our worked example demonstrated the potential impact of model specification methods by reanalyzing the data from one of our prior DBER investigations of equity (Van Dusen & Nissen, 2020) using each of the four model specification methods. Each of the model specification methods we did not recommend using created misspecified models, as our statistical analysis predicted they would. In this example, using statistical significance or variance explained to specify the model led to the identification of best models with biased results. The differences in the findings and the conclusions they support can have real-world consequences. For example, an administrator looking at the findings from the model specified using p-values (Model 8) would reasonably conclude not to allocate funds to support collaborative learning as it is not associated with improved student learning.

Education researchers should always reflect on their methods and how those methods influence findings. Critical examination of methods is especially warranted in investigations of equity where misleading findings can obscure and perpetuate harm caused by racism and sexism. In this investigation, we use a QuantCrit perspective to investigate the methods used in model specification. Based on our analysis, we recommend that DBER investigations of equity use information criterion to create intersectional models that minimize bias, and both minimize and accurately represent model uncertainty. Model specification, however, is only one step in a chain of quantitative methods that lead to findings. Each of these steps deserves similar scrutiny to that presented in this article. However, this is difficult for individual researchers interested in engaging in DBER investigations of equity when the field itself has not substantively examined many of these issues. Future investigations should examine how DBER investigations of equity are biased by statistical power, interpretations of model uncertainty, data filtering practices, handling of missing data, measures of equity, and data visualizations.

## 2.   9. References